\documentstyle[multicol,aps,prl,epsf,psfig]{revtex}
\begin{document}
\title{Nonequilibrium Phase Transitions in Models of Aggregation,
Adsorption and Dissociation}
\author{Satya N. Majumdar$^{1}$, Supriya Krishnamurthy$^{2}$ and Mustansir
Barma$^{1}$\\
{\small 1. \em Tata Institute of Fundamental Research, Homi Bhabha Road, 
Mumbai-400005, India}\\
{\small 2. \em PMMH, ESPCI, 10 Rue Vauquelin, 75231 Paris Cedex 05,
France}
}
\maketitle

\begin{abstract}
We study nonequilibrium phase transitions in a mass-aggregation model 
which allows for diffusion,
aggregation on contact, dissociation, adsorption and desorption of unit
masses. We analyse two limits explicitly. In the first case mass is
locally conserved whereas in the second case local conservation is
violated. In both cases the system undergoes a dynamical phase transition
in all dimensions.
In the first case, the steady state mass distribution decays exponentially
for large mass in one phase, and develops an infinite aggregate in
addition to a power-law mass decay in the other phase. In the second case,
the transition is similar except that the infinite aggregate is missing.

\end{abstract}

\begin{multicols}{2}
The steady state of a system in thermal equilibrium is described by
the Gibbs distribution. Phase transitions which occur in such equilibrium
systems
as one changes the external fields such as temperature or magnetic field
are by now well understood. On the other hand there is   
a wide variety of inherently {\em 
nonequilibrium} systems in nature 
whose steady states are not described by the Gibbs distribution, but are  
determined by the underlying microscopic dynamical processes and are
often hard to determine. Examples
include systems exhibiting self-organized criticality\cite{BTW}, several
reaction-diffusion systems\cite{ZGB}, fluctuating 
interfaces\cite{edw} and many others. As one changes the rates of
the underlying dynamical processes, the steady states of such systems
may undergo nonequilibrium phase transitions. As
compared to their equilibrium counterparts, these
nonequilibrium steady states and the transitions between them are much
less understood
due to the lack of a general framework. It is therefore important and 
necessary to study simple models amenable to analysis in order to
understand
the mechansims of such phase transitions.

Here we study the nonequilibrium phase transitions in an
important class of systems which involve microscopic processes of
diffusion and aggregation, dissociation, adsorption and desorption
of masses. These processes are ubiquitous in
nature, and arise in a variety of physical settings, for example, in
the formation of colloidal suspensions\cite{White} and polymer 
gels\cite{Ziff} on 
the one hand, and aerosols and clouds\cite{Fried} on the other. They
also enter in an important way in surface growth phenomena involving
island formation\cite{Lewis}. In this Letter, we introduce a simple
lattice model incorporating these microscopic processes and study the
nonequilibrium steady states and the transitions
between them both analytically within mean field theory and numerically
in one dimension.

Our lattice model, which evolves in continuous time, is defined as
follows.
For simplicity we define the model on a one-dimensional lattice with
periodic boundary conditions although generalizations to higher dimensions
are quite straightforward.
Beginning with a state in which the masses are placed
randomly, a site is chosen at random. Then one of the following events
can occur: 
\begin{enumerate}
\item Adsorption: With rate $q$, a unit mass is adsorbed at site $i$; thus
$m_i\to m_i+1$.
\item Desorption: With rate $p$, a unit mass desorbs from site $i$; thus
$m_i\to m_i-1$ provided $m_i\geq 1$.
\item Chipping (single-particle dissociation):
With rate $w$, a bit of the mass at the site
``chips'' off, {\it i.e.} provided $m_i\geq 1$, a single particle leaves
site $i$
and  moves with equal probability to one of the neighbouring
sites $i-1$ or $i+1$; thus $m_i\to m_i-1$ and $m_{i\pm 1}+1$.
\item Diffusion and Aggregation:
With rate $1$, the  mass $m_i$ at site $i$ moves either to
site $i-1$ or to site $i+1$.
If it moves to a site which already has some 
particles, then the total mass just adds up; thus $m_i\to 0$ and $
m_{i\pm 1}\to m_{i\pm 1}+m_i$.
\end{enumerate}

Note that we have assumed that both desorption and
diffusion rates are independent of the mass. In a more realistic situation
these rates would depend upon the mass. However, our aim here
is not to study this model in full generality, but rather to identify
the mechanism of a dynamical phase transition in the simplest possible
scenario involving these microscopic processes. Indeed we show below that
even within this simplest scenario, novel dynamical phase transitions
occur which are nontrivial yet amenable to analysis. 

Though the model can be studied in the full parameter space of all
four basic processes, for simplicity we restrict ourselves here to two
limiting cases: (i) $p=0$, $q=0$, i.e. only chipping, diffusion
and aggregation moves are allowed. In this limit, mass is locally 
conserved by the moves and we call this model the conserved-mass
aggregation
model (CMAM) (ii)
$w=0$, i.e. all moves except for chipping are allowed. In this case,
adsorption and desorption lead to violation of local mass
conservation. We call this the In-out model.
In this Letter, we analyse the CMAM model in some detail and
only outline the main results for the In-out model. 

Let us summarize our main results: (i) In the CMAM, single
particles are allowed to chip off from massive conglomerates.
This move corresponds to the physical process of single functional units 
breaking off from larger clusters in the polymerization problem. It leads
to a replenishment of the lower end of the mass spectrum, and competes
with the tendency of the coalescence process to produce more massive
aggregates. The result of this competition is that two types of
steady states are possible, and there is a dynamical phase
transition between the two. In one state, the steady state mass
distribution $P(m)$
decays exponentially, while
the other is more striking and interesting: $P(m)$ decays as a 
power law for large $m$ but in addition develops a delta function peak 
at $m=\infty$. Physically this means that an infinite aggregate forms 
that subsumes a finite fraction of the total mass, and coexists with 
smaller finite clusters whose mass distribution has a power law tail.
In the language of sol-gel transitions, the infinite aggregate is like the
gel while the smaller clusters form the sol. However, as opposed to the 
models of irreversible gelation where the sol disappears in the steady 
state, in our model the sol coexists with the gel even in the steady state.
Interestingly, the mechanism of formation of the infinite aggregate in
the steady state resembles Bose-Einstein condensation (BEC),
though the condensate (the infinite aggregate here) forms in real space
rather than momentum space as in conventional BEC. (ii) In the In-out
model too we find a phase transition in the steady state as the adsorption
($q$) and desorption ($p$) rates are varied.
In one phase (low values of $q$) $P(m)$ decays exponentially whereas
in the other phase (high $q$) it has a power law tail. This power law
phase is similar to that of the Takayasu model\cite{Takayasu} of particle
injection and aggregation.  

We first analyse the CMAM within the mean field approximation,
ignoring correlations in the occupancy of adjacent sites. Then we can
directly write down
equations for $ P(m,t)$, the probability that any site
has a mass $m$ at time $t$.
\\
\begin{eqnarray}
\frac{dP(m,t)} {dt} &=& -(1+w)[1+s(t)] P(m,t) 
+ w P(m+1,t)  \nonumber \\
&+&w s(t) P(m-1,t)+ P*P ;\;\;\; m \geq 1~~ \label{eq:mft1}\\
\frac{dP (0,t)} {dt} &=& - (1+w)s(t) P(0,t) + wP(1,t) + 
s(t) \label{eq:mft2}. 
\end{eqnarray} 
Here $ s(t) \equiv 1-P(0,t)$ is the probability that a site is
occupied by a mass and 
$P*P=\sum_{m^{\prime}=1}^{m}P(m^{\prime},t)P(m-m^{\prime},t)$ is a
convolution term that describes the coalescence of two masses.

The above equations enumerate all possible ways in  which the  mass 
at a site might change. The first term in Eq. (\ref{eq:mft1}) is
the ``loss'' term that accounts for the probability that a
mass $m$ might move as a whole or chip off to either of the neighbouring 
sites, or a 
mass from the neighbouring site might move or chip off to the site in 
consideration. The probability of occupation of the neighbouring site, 
$s(t) = \sum_{m=1} P(m,t)$, multiplies $P(m,t)$ within the mean-field
approximation where one neglects the spatial correlations in the
occupation probabilities of neighbouring sites. The remaining three terms
in Eq. (\ref{eq:mft1}) are the ``gain'' terms enumerating the number of
ways
that a site with mass $m^{\prime} \neq m$ can gain the deficit mass
$m -m^{\prime}$. The second equation Eq. (\ref{eq:mft2}) is a similar
enumeration of the possibilities for loss and gain of empty sites. 
Evidently, the mean field equations conserve the total mass.

To solve the equations, we compute the generating function, 
$Q(z,t) = \sum_{m=1}^{\infty} P(m,t)z^{m}$ from Eq. (\ref{eq:mft1}) and set 
$ \partial Q / \partial t =0$ in the steady state. We also need to use
Eq. (\ref {eq:mft2}) to write $P(1,t)$ in terms of $s(t)$. This gives
us a quadratic equation for $Q$ in the steady state.
Choosing the root that corresponds to
$Q(z=0) = 0$, we find
\\
\begin{eqnarray}
Q(z) &=&{{w+2s+ws}\over {2}}-{w\over {2z}}-{wsz\over {2}} 
\nonumber \\
&+& ws{(1-z)\over {2z}}\sqrt {(z-z_1)(z-z_2)}.  \label{eq:qsol}
\end{eqnarray}
where $z_{1,2}=(w+2\mp 2\sqrt {w+1})/ws$.
The value of the occupation probability $s$ is
fixed by mass conservation which implies that $\sum mP(m)=M/L\equiv \rho$.
Putting ${\partial}_zQ (z=1)=\rho$, the resulting relation between $\rho$ 
and $s$ is \begin{equation}
2\rho = w(1-s) - ws\sqrt {(z_1-1)(z_2-1)}~.
\label{eq:defq}
\end{equation}

The steady state probability distribution $P(m)$ is the coefficient of 
$z^m$ in $Q(z)$ and can be obtained from $Q(z)$ in Eq. (\ref{eq:qsol})
by evaluating the integral 
\begin{equation}
P(m) = {1\over {2\pi i}}\int_{C_o} \frac {Q(z)} {z^{ m+1}} dz
\label{eq:contour}
\end{equation}
over the contour $C_o$ encircling the origin. 
The singularities of the integrand govern the asymptotic behaviour of 
$P(m)$ for large $m$. Clearly the integrand has branch cuts at 
$z=z_{1,2}$. For fixed $w$, if one increases the density $\rho$, the
occupation probability $s$ also increases as evident from Eq. 
(\ref{eq:defq}).
As a result, both the roots $z_{1,2}$ start decreasing. As long as the
lower root $z_1$ is greater than $1$, Eq. (\ref{eq:defq}) is well defined
and the analysis of the contour integration around the branch cut
$z=z_1$, yields for large $m$,
\begin{equation}
P(m) \sim e^{-m/{m^{*}}}/m^{3/2} ~,
\end{equation}
where the characteristic mass, $m^{*}=1/{\log (z_1)}$ and diverges as
$\sim (s_c-s)^{-1}$ as $s$ approaches $s_c =(w+2-2\sqrt {w+1})/w$. 
$s_c$ is the critical value of $s$ at which $z_1=1$. This exponentially 
decaying mass distribution is the signature of ``disordered" phase which 
occurs for $s<s_c$ or equivalently from Eq. (\ref{eq:defq}) for
$\rho < {\rho}_c={\sqrt {w+1}} -1$.

When $\rho={\rho}_c$, we have $z_1=1$, and analysis of the contour around 
$z=z_1=1$ yields a power law decay of $P(m)$,
\begin{equation}
P(m)\sim m^{-5/2}.
\end{equation}
As $\rho$ is increased further beyond ${\rho}_c$, $s$ cannot increase
any more because if it does so, the root $z_1$ would be less than $1$ 
(while the other root $z_2$ is still bigger than $1$) and
Eq. (\ref{eq:defq}) would be undefined. The only possibility is that $s$
sticks to its critical value $s_c$ or equivalently the lower root $z_1$
sticks to $1$. Physically this implies that adding more particles   
does not change the occupation probability of sites. This can happen only 
if all the additional particles (as $\rho$ is increased) aggregate on a 
vanishing fraction of sites, thus
not contributing to the occupation of the others. Hence in this 
``infinite-aggregate" 
phase $P(m)$ has an infinite-mass aggregate, in addition to the power law
decay $m^{-5/2}$. Concomitantly Eq. (\ref{eq:defq}) ceases to hold, and 
the relation now becomes 
\begin{equation}
\rho = {w\over {2}}(1-s_c) + \rho_{\infty}
\end{equation}
where $\rho_{\infty}$ is the fraction of the mass in the infinite aggregate.
The mechanism of formation of the aggregate is reminiscent of Bose 
Einstein condensation. In that case, for temperatures in which a macroscopic
condensate exists, particles added to the system do not contribute to the 
occupation of the excited states; they only add to the condensate, as 
they do to the infinite aggregate here.

Thus the mean field phase diagram (see inset of Fig. 1) of the system
consists of two phases,
``Exponential" and ``Aggregate", which are separated by the
phase boundary, $\rho_c={\sqrt {w+1}}-1$. While this
phase diagram remains qualitatively the same even in $1$-d, 
the exponents characterizing the power laws are different from
their mean field values (see Fig. 1).

\begin{figure}
\begin{center}
\leavevmode
\psfig{figure=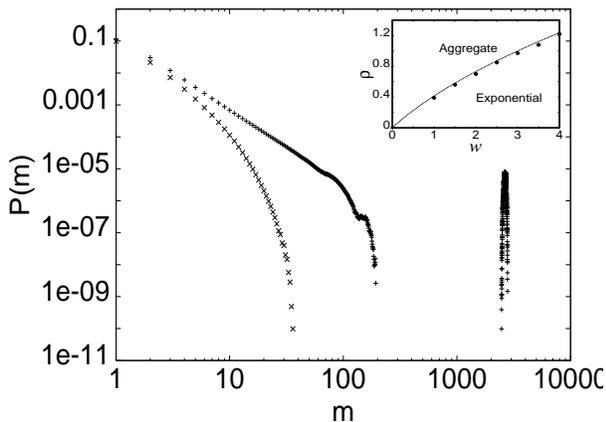,width=8cm,height=6cm,angle=0}
\end{center}
\narrowtext
\caption{log-log plot of $P(m)$ vs. $m$ for the CMAM, for
$\rho <, =, > {\rho}_c$ on a lattice with $L=1000$. Inset: Phase diagram.
The solid line and the points 
indicate the phase boundary within mean field theory and
$1$-d simulation respectively.}
\end{figure}

We have studied this model using Monte Carlo simulations on a
one-dimensional lattice. Although we present results here for a 
relatively small  size lattice, $L=1024$, we have checked our results for
larger sizes as well. We confirmed that all
the qualitative predictions of the mean-field theory remain true, by
calculating $P(m)$ numerically in the steady state.  Figure 1 
displays two numerically obtained plots of
$P(m)$. The existence of both the ``Exponential" (denoted by $\times$)
for $\rho<{\rho}_c$
and the ``Aggregate" phase (denoted by $+ $) for $\rho>{\rho}_c$ is
confirmed. 
In particular, the second curve shows clear evidence of a power-law behaviour
of the distribution, which is cut off by finite-size effects, and
for an `infinite' aggregate beyond. We confirmed that the mass $M_{agg}$
in this aggregate grows linearly with the size, and that the
spread $\delta M_{agg}$ grows sublinearly, implying that the ratio
$\delta M_{agg}/M_{agg}$ approaches zero in the thermodynamic limit.
The exponent $ \tau_{CMAM} $ which characterizes the finite-mass fragment
power law decay for $\rho>{\rho}_c$ is numerically found to be $ 2.33\pm
.02$ and remains the same at the critical point $\rho=\rho_c$.

We note that in conserved-aggregation models studied
earlier within mean field theory\cite{Vigil,Krapiv}, the steady state mass
distribution also changed from an exponential distribution to a power law
as the density was increased beyond a critical value. However, the
existence of the striking infinite aggregate in the steady state for
$\rho>{\rho}_c$ was not identified earlier.

We next study the steady state phase diagram of the In-out model
in the $q$-$p$ plane. In this model, mass is not locally conserved. 
The mass per site $M(t)$ evidently obeys the exact equation
\begin{equation}
\frac{dM}{dt}=q-ps(t)
\label{eq:mass}
\end{equation}
where $ s(t)$ is the fraction of sites occupied by a mass $m_i \geq 1$.
In the steady state, let the mean value of $s(t)$ be $s$. If
$q/p$ is low, $s$ adjusts to make $q-ps$ vanish, and the mean mass
reaches a time-independent value $M$. This defines the finite-mass phase.
As we will see below, as $q/p$ increases beyond a critical value, $s$
never catches up with $q/p$ and reaches a steady state value which
is less than $q/p$; in this phase, $M$ increases
linearly in time while $P(m,t)\sim m^{-\tau_{T}}f(mt^{-x})$ which
in the long time limit converges to a time-independent form, decaying as a
power law with exponent $\tau_{T}$, even though the moments of
this distribution diverge as time increases to infinity. We call this
the growing-mass or the Takayasu phase. 
In fact, for $p=0$ the In-out model reduces exactly to that of the
Takayasu model (TM) of injection and aggregation of
masses\cite{Takayasu}
which has found widespread applications ranging from river
models\cite{river}
to granular systems\cite{SNC}. Indeed what we find here is that the
growing mass phase of the TM at $p=0$ persists up to a nonzero critical
value $p_c(q)$ for a given $q$, while for $p>p_c(q)$ the mass stops
growing and $P(m)$ decays exponentially for large $m$ in the steady state.

The mean field analysis of the In-out model is similar to that of the
CMAM model though a little bit trickier. We defer the details for a
future publication\cite{KMB} and only outline the results here.
We find that the critical line 
$p_c(q)=q+2{\sqrt {q}}$ separates two phases in the $q$-$p$ plane. For
$p>p_c$, $P(m)\sim m^{-3/2}\exp (-m/m^{*})$ for large $m$. For $p=p_c$,
$P(m)\sim m^{-5/2}$ and for $p<p_c$, $P(m)\sim m^{-3/2}$ for large $m$.
For a fixed $q$, the steady state occupation density $s(p,q)$ develops
an interesting cusp as $p$ crosses $p_c(q)$. For example, at $q=1$,
where $p_c=3$, $s(p)=1/p$ for $p>3$ as follows simply from Eq. (9);
but for $p<3$, the determination of $s(p)$ is nontrivial\cite{KMB} and
is given by the
positive root of the cubic equation,
$16ps^3+(8p^2+4p-25)s^2+(p^3-11p^2-43p-25)s-p^3+2p^2+17p+25=0$.

The qualitative predictions of mean field theory -- the existence of
a power-law (Takayasu) phase ($P(m) \sim m^{-\tau_T}$)
and a phase with exponential mass
distribution, with a different critical behaviour at the transition 
($P(m) \sim m^{-\tau_c}$) -- are found to hold in 1-d as well. The
Takayasu exponent $\tau_T$ is known exactly to be $4/3$ in $1$-d
and $3/2$ within mean field theory\cite{Takayasu}. 
Figure 2 shows the results of numerical simulations in $1$-d for the phase
diagram and the decay 
of the mass distribution in the two phases and at the transition point.
The values obtained, $\tau_T=4/3$ and $\tau_c \simeq 1.833$,  
are quite different from their mean-field values, $\tau_T=3/2$ and
$\tau_c =5/2$, reflecting the effects of correlations between masses at
different sites. 

\begin{figure}
\psfig{figure=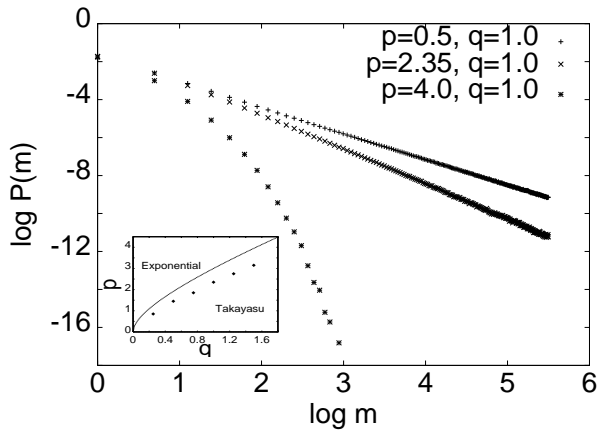,width=8cm,angle=0}
\narrowtext
\caption{log-log plot of $P(m)$ vs. $m$ for the In-out model,
for $p<,= , > p_c$. Inset: Phase diagram. The solid line and
the points indicate the phase boundary
within mean field theory and $1$-d simulation respectively.}
\end{figure}

We may reinterpret the configuration of masses in the In-out model as an
interface profile on regarding $m_i$ as a local height variable. 
While the model may have some unphysical features in the context of an
interface due to the columns of masses moving as a whole, the analogy
helps however to understand physically the nature
of the transition in the In-out model. In the interface language
this corresponds to a wetting transition; the key factor responsible for
the occurrence of the smooth phase is a substrate, implicit in the
constraint $m_i\geq 0$ in the In-out model. The wet phase is identified
with a growing mass phase, which has a rough profile, with exact 
roughness exponent $\chi_{T}=5/2$\cite{KMB} in $1$-d. Since $\chi_T>1$,
the
interface in the wet phase is not self-affine. Recently a nonequilibrium
wetting transition was also observed in an interface
model\cite{Hinrichsen} where the interface in the wet phase is self-affine
due to surface tension effects which are absent in our model.
Interestingly, however, in our model the substrate is able to induce a
self-affine interface at the critical point with roughness exponent
$\chi =1/3$ within mean field theory and $\chi\approx 0.7$ in
$1$-d\cite{KMB}, despite the anomalously large roughness of the wet phase.
 
We thank Deepak Dhar for useful discussions and S. Cueille and S. Redner
for pointing out references \cite{Vigil} and \cite{Krapiv}
to us.

\end{multicols}

\end{document}